\documentclass{easychair}
\usepackage{latexsym}

\newtheorem{claim}{Claim}
\newtheorem{example}{Example}

\newcommand{\nikolaj}[1]{} 

\titlerunning{HOPA as SMT with Algebraic Data-types}
\authorrunning{Bj{\o}rner, McMillan, Rybalchenko}

\author{
\\Nikolaj Bj{\o}rner\\
Microsoft Research\\
\and
\\Ken McMillan\\
Microsoft Research\\
\and
\\Andrey Rybalchenko\\
Microsoft Research \& Technische Universit{\"a}t M{\"u}nchen
}

\title{Higher-order Program Verification as Satisfiability Modulo Theories with Algebraic Data-types}

\begin{document}

\maketitle

\begin{abstract}
We report on work in progress on automatic procedures for
proving properties of programs written in higher-order functional 
languages. Our approach encodes higher-order programs directly as 
first-order SMT problems over Horn clauses.
It is straight-forward to reduce Hoare-style verification
of first-order programs into satisfiability of Horn clauses.
The presence of closures offers several challenges:
relatively complete proof systems have to account for closures;
and in practice, the effectiveness of search procedures
depend on encoding strategies and capabilities of underlying solvers.
We here use algebraic data-types to encode closures and 
rely on solvers that support algebraic data-types. 
The viability of the approach is examined using examples
from~\cite{DBLP:conf/popl/UnnoTK13}.
\end{abstract}

\section{Background}
Automatic verification of programs in higher-order languages has received 
significant attention in recent years. The Liquid Type systems have been used
successfully on a large set of challenges from 
Ocaml~\cite{DBLP:conf/pldi/RondonKJ08} and 
Haskell~\cite{DBLP:conf/esop/VazouRJ13}.
Liquid types rely checking program properties as 
type checking of refinement types.
Not all refinement types need to be provided fully, 
instead users can supply a space
of templates that the type inference engine instantiates and refines 
the template invariants 
(using the Houdini approach~\cite{DBLP:conf/fm/FlanaganL01}).
A very different approach is taken in 
HALO~\cite{DBLP:conf/popl/VytiniotisJCR13} where the denotational 
semantics of Haskell, using types that include error states,
is encoded using quantified equalities. HALO relies on quantifier
instantiation engines and finite model building capabilities to discharge 
correctness proof conditions. 
Unno et.al.~\cite{DBLP:conf/popl/UnnoTK13} develop a custom engine 
that realizes proof rules for Hoare logic for programs with procedure
parameters~\cite{DBLP:journals/iandc/GermanCH89}. 
There are several other systems that use first-order/SMT technologies for
establishing properties of higher-order languages. They typically
rely on user-annotated refinement types. The executable subset of the 
PVS specification language uses decision procedures for discharging
type checking conditions and the F$\star$ system includes 
verification condition generation 
from a higher-order language with higher-order refinement types.

These systems have in common that they rely on SMT solvers, 
CVC4, Yices and Z3. We are here examining whether
program verification (in the presence of closures) can be 
reduced directly to SMT solving. 

\section{Program Verification as Horn Clause Satisfiability}

Consider the McCarthy function:
\begin{verbatim}
let mc x = if x > 100 then x - 10 else mc (mc (x + 11))
assert x <= 101 => mc(x) = 91
\end{verbatim}

We can check \emph{partial correctness} of \texttt{mc} by representing the summary as a binary relation
between input and outputs, and describing the function and assertion 
equivalently as a conjunction of Horn clauses:
\begin{eqnarray*}
\forall x \ . \ x > 100 \ \rightarrow \ \mathtt{mc}(x,x-10) \\
\forall x,y,z \ . \ x \leq 100 \;\land\; \mathtt{mc}(x+11, y) \;\land\; \mathtt{mc}(y,z) \ \rightarrow \ \mathtt{mc}(x,y) \\
\forall x,y \ . \ x \leq 101\; \land\; \mathtt{mc}(x,y) \ \rightarrow \ y = 91
\end{eqnarray*}
These Horn clauses are satisfiable if and only if the original partial
correctness assertion holds.
A satisfying assignment to the Horn clauses (found by Z3) is:
\begin{eqnarray*}
mc(x,y) & \equiv & (y \leq x - 10 \; \lor y \leq 91) \land y \geq 91 \land (x \leq y + 10)
\end{eqnarray*}
The assignment is an \emph{inductive} summary for $\mathtt{mc}$.

This example illustrates a central claim made in~\cite{SMT2012}
and~\cite{HSF}:
\begin{claim} Satisfiability of Horn clauses modulo theories 
is an adequate basis for program correctness.
\end{claim}
Synthesis of ranking functions can also be encoded into 
Horn clauses~\cite{HSF}, so Horn clause solving extends 
also to establishing \emph{total} correctness.

The challenge is of course to \emph{solve} Horn clauses and a number of tools
are being developed for solving Horn clauses at scale. These include
the Duality~\cite{Duality}, 
HSF~\cite{HSF}, 
Eldarica~\cite{RuemmerETAL13DisjunctiveInterpolantsHornClauseVerification},
PDR implementation in Z3~\cite{PDR}, and
SPACER~\cite{Spacer}. 
The Horn clause format is also a convenient interchange 
format for symbolic software verification benchmarks.
We have collected benchmarks 
from symbolic software model checking
as Horn clauses in an online repository~\footnote{
\url{https://svn.sosy-lab.org/software/sv-benchmarks/trunk/clauses/}}.
There are currently around 10,000 benchmarks from various sources.
There are several possible strategies for encoding model checking problems
into Horn clauses and consequently some benchmarks use
different encodings of the same problems. 

The repository also includes problems from the tools for liquid types. 
The liquid type system leverages the Hindley Milner type system for extracting
Horn clauses for closures.


\section{Closures as Algebraic Data-types}
We will here consider a direct encoding of problems from 
a simply typed higher-order programming language into Horn clauses. 
The basic idea is to encode functions as relations and higher-order 
closures as algebraic data-types. The Horn clauses encode an interpreter, 
suitably specialized to the programs that are analyzed.
We use examples from~\cite{DBLP:conf/popl/UnnoTK13} 
to illustrate the approach.

\newcommand{\constr}[1]{\underline{\mathtt{#1}}}

To warm up, consider the program
\begin{verbatim}
let f x y = assert (not (x() > 0 && y() < 0))
let h x y = x
let g n = f (h n) (h n)
\end{verbatim}
The curried function \texttt{h} is partially applied to the same integer $n$, 
and \texttt{f} is applied to two functions with signature $\mathtt{unit}\rightarrow\mathtt{int}$.
This is the only closure type used in this program, so we introduce algebraic data-types
to encode the possible closures that are used in the program:
\begin{eqnarray*}
\mathit{clo} & ::= & \constr{h}\ \mathtt{int} \\
\mathit{unit} & ::= & \constr{unit}
\end{eqnarray*}
The closure $\mathtt{clo}$ has a single constructor $\constr{h}$ that takes an integer.
It is evaluated to arguments of type $\mathit{unit}$.
We assign meaning to the closure by defining an evaluator. 
A canonical evaluator can be formulated as a relation over 
(1) a closure, (2) the argument of the closure, (3) the output value,
and (4) a flag to indicate successful termination of the evaluation.
For our example, there is only a single relevant rule for the evaluator;
it reduces to evaluating the function $h$
\begin{eqnarray*}
\forall x, r, \mathit{ok} \ . \ \mathit{h}(x,\constr{unit}, r, \mathit{ok}) & \rightarrow & \mathit{Ev}(\constr{h}(x),\constr{unit}, r, \mathit{ok})
\end{eqnarray*}
The other functions translate in a straight-forward way to Horn clauses:
\begin{eqnarray*}
	\left(
	\begin{array}{ll}
	&	\mathit{Ev}(x,\constr{unit},r_1,\mathit{ok}_1) \\
	\; \land\; & \mathit{Ev}(y,\constr{unit},r_2,\mathit{ok}_2) \\
	\; \land\; & (\mathit{ok} \equiv \neg(r_1 > 0 \land r_2 < 0) \land \mathit{ok}_1 \land \mathit{ok}_2)
	\end{array}
	\right)
		& \rightarrow & f(x,y,\constr{unit},\mathit{ok}) \\
& & h(x,y,x,\constr{true}) \\
f(\constr{h}(n),\constr{h}(n),r,\mathit{ok}) & \rightarrow & g(n,r,\mathit{ok}) \\
g(n,r,\constr{false}) & \rightarrow & \constr{false}
\end{eqnarray*}
where $x, y, r, r_1, r_2, \mathit{ok}, \mathit{ok}_1, \mathit{ok}_2, n$ are variables.

We used this example to sketch a systematic encoding from functional programs to Horn clauses.
It already contains some shortcuts (alternatively, we could have considered defining a generic interpreter). 
It can still be simplified by specializing the program to the correctness assertion and removing arguments.
Since $\constr{h}(x)$ is the \emph{only} closure passed to $\mathit{Ev}$ it can furthermore be removed entirely.
The simpler, equi-satisfiable set of clauses is:
\begin{eqnarray*}
\mathit{h}(x, \constr{unit}, r) & \rightarrow & \mathit{Ev}(x, r) \\
	\mathit{Ev}(x,r_1) 
	\; \land\; \mathit{Ev}(y,r_2) 
	\; \land\;  r_1 > 0 \land r_2 < 0	& \rightarrow & f(x,y) \\
& & h(x,y,x) \\
f(n, n) & \rightarrow & g(n,r) \\
g(n,r) & \rightarrow & \constr{false}
\end{eqnarray*}
where $x, y, r_1, r_2, n, r$ are variables.
The Horn clauses are non-recursive and Z3's engines for Horn clauses can easily establish satisfiability
(e.g., that the assertion holds).

A significantly more challenging example is suggested in~\cite{DBLP:conf/popl/UnnoTK13} 
to illustrate their method (that comprises of adding
extra parameters to closures).

\begin{verbatim}
let app1 f g = if * then app1 (succ f) g else g f
let app2 i f = f i 
let succ f x = f (x + 1)
let check x y = assert (x <= y)
let main i = app1 (check i) (app2 i)
\end{verbatim}

We use the shortcuts used in the previous example to encode the recursive functions in an economical way.
The example uses two closures corresponding to the types $\constr{check}(i): \mathtt{int} \rightarrow \mathtt{unit}$
and $\constr{app2}(i): (\mathtt{int} \rightarrow \mathtt{int}) \rightarrow \mathtt{unit}$.
\begin{eqnarray*}
\mathit{clo}_1 & ::= & \constr{app2}\ \mathtt{int} \\
\mathit{clo}_2 & ::= & \constr{check}\ \mathtt{int} \; | \; \constr{succ}\ \mathit{clo}_2
\end{eqnarray*}
Similar to the previous example, the closure $\mathit{clo}_1$ is superfluous and does not need to appear in the generated Horn clauses.
The second closure is recursive and therefore essential, 
and we include this in the corresponding Horn clauses.
Let $x, y, f, i$ be variables, then safety of the example program
is equivalent to satisfiability of the following set of clauses:
\begin{eqnarray*}
\mathit{app}_1(\constr{succ}(f), i) & \rightarrow & \mathit{app}_1(f, i) \\
\mathit{Ev}(f, i) & \rightarrow & \mathit{app}_1(f, i) \\
\mathit{Ev}(f, x + 1)             & \rightarrow & \mathit{succ}(f, x) \\
x > y & \rightarrow & \mathit{check}(x,y) \\
\mathit{app}_1(\constr{check}(i), i) & \rightarrow & \mathit{main}(i) \\
\mathit{main}(i) & \rightarrow & \constr{false} \\
\mathit{succ}(f, x) & \rightarrow & \mathit{Ev}(\constr{succ}(f),x) \\
\mathit{check}(x,y) & \rightarrow & \mathit{Ev}(\constr{check}(x),y) 
\end{eqnarray*}

Z3 accepts recursive Horn clauses with algebraic data-types as input, 
but is unable to establish satisfiability of these clauses directly.
In a nutshell, it lacks the ability to synthesize properties that 
select leaves in algebraic data-types.
We will here examine a couple of approaches that could be used to
solve such problems.

\subsection{In-lining by resolution}
Our first approach is to inline definitions by resolving
Horn clauses and merge predicates. The result is given below
where $\mathit{app}_1$ and $\mathit{Ev}$ are merged into $\mathit{Ev}$.

\begin{eqnarray*}
\mathit{Ev}(\constr{succ}(f), i) & \rightarrow & 
\mathit{Ev}(f, i) \\
\mathit{Ev}(f, i) & \rightarrow & 
\mathit{Ev}(f, i) \\
\mathit{Ev}(\constr{check}(i), i) & \rightarrow & \constr{false} \\
\mathit{Ev}(f, x+1) & \rightarrow & 
\mathit{Ev}(\constr{succ}(f),x) \\
x > y & \rightarrow & 
\mathit{Ev}(\constr{check}(x),y) 
\end{eqnarray*}

The second clause is a tautology and the two clauses that 
contain $\constr{succ}(f)$ can be resolved (and then removed, 
assuming a suitably strong, but not entirely unrealistic, 
redundancy elimination mechanism)
leaving an equi-satisfiable set of Horn clauses:

\begin{eqnarray*}
\mathit{Ev}(\constr{check}(i), i) & \rightarrow & \constr{false} \\
\mathit{Ev}(f, x+1) & \rightarrow & \mathit{Ev}(f,x) \\
x > y & \rightarrow & \mathit{Ev}(\constr{check}(x),y) 
\end{eqnarray*}
Z3 can immediately establish satisfiability of this set of clauses.

\subsection{Quantified abstraction}

In~\cite{SAS} we consider synthesis of intermediary assertions
using quantified invariants for array manipulating programs.
Our approach from~\cite{SAS} is to search over solutions
over a template space of quantified invariants. The idea
is to supply additional arguments to recursive predicates
and bind these in quantifiers.
Using the running example, the modified problem is:
\begin{eqnarray*}
(\forall \vec{z}\ .\ \mathit{Ev}(\vec{z},\constr{succ}(f), i)) & \rightarrow & 
(\forall \vec{z}\ .\ \mathit{Ev}(\vec{z}, f, i)) \\
(\forall \vec{z}\ .\ \mathit{Ev}(\vec{z}, \constr{check}(i), i)) & \rightarrow & \constr{false} \\
(\forall \vec{z}\ . \ \mathit{Ev}(\vec{z}, f, x+1)) & \rightarrow & 
(\forall \vec{z}\ . \ \mathit{Ev}(\vec{z}, \constr{succ}(f),x)) \\
x > y & \rightarrow & 
(\forall \vec{z}\ . \ \mathit{Ev}(\vec{z}, \constr{check}(x),y)) 
\end{eqnarray*}
where $\vec{z}$ is a tuple of integer variables. 
The number of variables in $\vec{z}$ is a parameter to the abstraction
procedure.
The instantiation heuristic suggested in~\cite{SAS} is very simple
and will not produce useful instantiations for this case. 
Instantiation heuristics that are more powerful for arithmetic
are used in~\cite{ESolving}, and of course the template 
method of~\cite{DBLP:conf/popl/UnnoTK13}. 

Let us for the sake of illustrating the idea (but not the practicality)
pretend an instantiation procedure produces the following set of clauses:
\begin{eqnarray*}
\mathit{Ev}(u, v,\constr{succ}(f), i) & \rightarrow & 
\mathit{Ev}(u, v, f, i) \\
\mathit{Ev}(i, i, \constr{check}(i), i) & \rightarrow & \constr{false} \\
\mathit{Ev}(u, v, f, x + 1) & \rightarrow & 
\mathit{Ev}(u, v + 1, \constr{succ}(f),x) \\
x > y & \rightarrow & 
\mathit{Ev}(u, v, \constr{check}(x),y) 
\end{eqnarray*}

With some struggle, Z3 can produce the inductive invariant:
\begin{eqnarray*}
\mathit{Ev}(u, v, \_, i) & \equiv & 2\cdot u = v + i
\end{eqnarray*}
More specifically, the invariant is obtained by analyzing
the program obtained by reversing the rules
(corresponding to a Magic set transformation)
\begin{eqnarray*}
\mathit{Ev}(u, v,\constr{succ}(f), i) & \leftarrow & 
\mathit{Ev}(u, v, f, i) \\
\mathit{Ev}(i, i, \constr{check}(i), i) & \leftarrow & \constr{false} \\
\mathit{Ev}(u, v, f, x + 1) & \leftarrow & 
\mathit{Ev}(u, v + 1, \constr{succ}(f),x) \\
x > y & \leftarrow & 
\mathit{Ev}(u, v, \constr{check}(x),y) 
\end{eqnarray*}
Karr's algorithm~\cite{DBLP:journals/acta/Karr76} 
for computing all affine relations for
Horn clauses produces the useful congruence.

\section{Summary}
We have examined using algebraic data-types to encode closures directly 
in Horn clauses. Algebraic data-types (Herbrand terms) can be used to 
provide a simple encoding of higher-order programs
into Horn clauses. Existing Horn clause solvers currently emphasize 
solving clauses over integers, reals, and more lately arrays. Solving
Horn clauses with algebraic data-types (Herbrand terms) pose a set of 
new challenges for such solvers, but we have illustrated a couple of 
independent approaches that can be used to solve such Horn clauses
directly.


\bibliographystyle{plain}
\bibliography{refs}

\begin{thebibliography}{10}

\bibitem{ESolving}
Tewodros Beyene, Corneliu Popeea, and Andrey Rybalchenko.
\newblock {Solving Existentially Quantified Horn Clauses}.
\newblock In {\em {CAV}}, 2013.

\bibitem{SAS}
Nikolaj Bj{\o}rner, Ken McMillan, and Andrey Rybalchenko.
\newblock {On Solving Quantified Horn Clauses}.
\newblock In {\em SAS}, 2013.

\bibitem{SMT2012}
Nikolaj Bj{\o}rner, Kenneth~L. McMillan, and Andrey Rybalchenko.
\newblock Program verification as {Satisfiability} {Modulo} {Theories}.
\newblock In {\em SMT}, 2012.

\bibitem{DBLP:conf/fm/FlanaganL01}
Cormac Flanagan and K.~Rustan~M. Leino.
\newblock Houdini, an annotation assistant for esc/java.
\newblock In Jos{\'e}~Nuno Oliveira and Pamela Zave, editors, {\em FME}, volume
  2021 of {\em Lecture Notes in Computer Science}, pages 500--517. Springer,
  2001.

\bibitem{DBLP:journals/iandc/GermanCH89}
Steven~M. German, Edmund~M. Clarke, and Joseph~Y. Halpern.
\newblock Reasoning about procedures as parameters in the language l4.
\newblock {\em Inf. Comput.}, 83(3):265--359, 1989.

\bibitem{DBLP:conf/popl/2013}
Roberto Giacobazzi and Radhia Cousot, editors.
\newblock {\em The 40th Annual ACM SIGPLAN-SIGACT Symposium on Principles of
  Programming Languages, POPL '13, Rome, Italy - January 23 - 25, 2013}. ACM,
  2013.

\bibitem{HSF}
Sergey Grebenshchikov, Nuno~P. Lopes, Corneliu Popeea, and Andrey Rybalchenko.
\newblock Synthesizing software verifiers from proof rules.
\newblock In {\em PLDI}, 2012.

\bibitem{PDR}
Kry{\v{s}}tof Hoder and Nikolaj Bj{\o}rner.
\newblock Generalized property directed reachability.
\newblock In {\em SAT}, 2012.

\bibitem{DBLP:journals/acta/Karr76}
Michael Karr.
\newblock Affine relationships among variables of a program.
\newblock {\em Acta Inf.}, 6:133--151, 1976.

\bibitem{Spacer}
Anvesh Komuravelli, Arie Gurfinkel, Sagar Chaki, and Edmund Clarke.
\newblock {Automatic Abstraction in SMT-Based Unbounded Software Model
  Checking}.
\newblock In {\em CAV}, 2013.

\bibitem{Duality}
Kenneth~L. McMillan and Andrey Rybalchenko.
\newblock Computing relational fixed points using interpolation.
\newblock Technical Report MSR-TR-2013-6, Microsoft Research, 2013.
\newblock {\tt http://research.\-microsoft.\-com/\-apps/\-pubs/\-?id=180055}.

\bibitem{DBLP:conf/pldi/RondonKJ08}
Patrick~Maxim Rondon, Ming Kawaguchi, and Ranjit Jhala.
\newblock Liquid types.
\newblock In Rajiv Gupta and Saman~P. Amarasinghe, editors, {\em PLDI}, pages
  159--169. ACM, 2008.

\bibitem{RuemmerETAL13DisjunctiveInterpolantsHornClauseVerification}
Philipp R\"ummer, Hossein Hojjat, and Viktor Kuncak.
\newblock Disjunctive interpolants for horn-clause verification.
\newblock In {\em CAV}, 2013.

\bibitem{DBLP:conf/popl/UnnoTK13}
Hiroshi Unno, Tachio Terauchi, and Naoki Kobayashi.
\newblock Automating relatively complete verification of higher-order
  functional programs.
\newblock In Giacobazzi and Cousot \cite{DBLP:conf/popl/2013}, pages 75--86.

\bibitem{DBLP:conf/esop/VazouRJ13}
Niki Vazou, Patrick~Maxim Rondon, and Ranjit Jhala.
\newblock Abstract refinement types.
\newblock In Matthias Felleisen and Philippa Gardner, editors, {\em ESOP},
  volume 7792 of {\em Lecture Notes in Computer Science}, pages 209--228.
  Springer, 2013.

\bibitem{DBLP:conf/popl/VytiniotisJCR13}
Dimitrios Vytiniotis, Simon L.~Peyton Jones, Koen Claessen, and Dan Ros{\'e}n.
\newblock Halo: haskell to logic through denotational semantics.
\newblock In Giacobazzi and Cousot \cite{DBLP:conf/popl/2013}, pages 431--442.

\end{thebibliography}

\end{document}